\documentclass[10pt]{iopart}

\usepackage{cite}
\usepackage{units}
\usepackage{tikz}
\usetikzlibrary{patterns}

\newcommand{\text}[1]{\mathrm{#1}}

\def\ioptwocol{\setlength\hoffset{-0.5in}\setlength\voffset{-0.5in}\setlength\textwidth{6.75in}
\setlength\columnsep{0.2in}\setlength\textheight{9.25in}\mathindent=0in\twocolumn}

\begin{document}
\ioptwocol

\title{An impulse to the ground to end rolling with slipping}

\author{Gerrit Ansmann}

\address{Institute for Biological Physics, University of Cologne, Z\"ulpicher Stra\ss{}e 47a, 50674 K\"oln, Germany}

\begin{abstract}
Several scenarios used in teaching feature a rolling motion with slipping that transitions to one without through friction with the ground.
We summarise these transitions by introducing an unknown impulse that is transferred to the ground.
Accounting for this in the conservation of angular and linear momentum, we can deduce the final state of these scenarios in a compact manner and without requiring details of the friction.
Contrasting this technique with an explicit calculation of the friction illustrates how collisions and conservation laws allow to solve problems involving complex interactions by summarising them.
We exemplify our technique with three scenarios:
a moving ball starting to roll, a turning wheel being released on the ground, and a monowheel breaking.
\end{abstract}

\vspace{2pc}

\section{Introduction}

Collisions as a physical concept use conservation laws or the transfer of momentum, respectively, to describe the outcome of a short and complex interaction whose details are otherwise infeasible or impossible to treat.
Not only are classical (linear) collisions ubiquitous in many fields, e.g., when treating particle interactions, but the concept can also be generalised to other kinds of kinetics and conservation laws, e.g., rotational collisions~\cite{Bougie2019, Danna2021}.
When teaching a simplifying approach like collisions, it is desirable to contrast it with the unravelled alternative, to demonstrate how and that it works as well as to deepen the understanding.
This is commonly done with energy conservation, where work against a potential can be calculated explicitly in many cases.
An example involving an elastic collision is the slingshot manoeuvre~\cite[Problem 107]{tipler2008}, which can be treated as a collision but also be expanded to a one-body problem.
However, there is a dearth of examples where energy conservation does not hold, one reason being that most inelastic collisions can only be unpacked into a complex rheological problem that goes far beyond the scope of an introduction to mechanics.

We here propose to apply a collision-like framework to problems involving rolling with slipping (RWS)~\cite{hierrezuelo1995, desousa1997, pinto2001, tipler2008, deambrosis2015, onorato2015, maidana2016, suarez2020, ozkanlar2020, kajiyama2021}.
These feature a \emph{roller,} i.e., a rigid solid with a shape allowing it to roll on a plane.
The roller is oriented such that it can roll, but its linear velocity~\(v\) and angular velocity~\(\omega\) do not meet the rolling condition \(v = \omega r\), with \(r\)~being the roller's radius.
In consequence, kinetic friction acts on the roller until the rolling condition is fulfilled and it rolls without slipping.
Recently, RWS experiments have gained attention as digital video-tracking techniques have made them accessible for teaching \cite{deambrosis2015, onorato2015, maidana2016, suarez2020, kajiyama2021}.
RWS problems are usually solved analytically by determining the
friction acting on the roller and calculating its simultaneous effects on \(v\) and~\(\omega\) (via the torque) until they meet the rolling condition~\cite{hierrezuelo1995, desousa1997, pinto2001, tipler2008, maidana2016, suarez2020, ozkanlar2020, kajiyama2021}.
While this detailed procedure has its didactic and practical merits, e.g., for determining the duration of the slipping phase, we here demonstrate a technique for determining the roller's final state directly without knowing the details of the friction:
We introduce an unknown impulse that is transferred to the ground and summarises the friction.
Inserting this impulse and the unknown final velocity (or angular velocity) in the equations for the conservation of momentum and angular momentum, we obtain two equations, which we can solve for the final velocity.

We propose to contrast both approaches to RWS problems to demonstrate how collisions summarise complex interactions.
Moreover, we show that the proposed technique allows to solve more difficult problems.
In the following, we exemplify our approach on three RWS problems and compare it to alternatives.
In the appendix, we provide exercises that illustrate and employ the proposed approach.

\section{Scenario A: Pushed roller} \label{Bowling}

\begin{figure*}
	\begin{tikzpicture}[scale=0.685,line cap=round]
	\fill[black!10!white]
		(0,0) rectangle (25,-0.5);
	\draw[thick] (0,0) -- (25,0);
	
	\draw[<->] (1,0) -- ++(0,2) node[midway,left] {\(r\)};
	
	\begin{scope}[shift={(4,0)}]
		\node at (0,5) {initial state};
		\filldraw[fill=black!20!white,thick] (0,0) arc(-90:270:2);
		\draw[->] (0,2) -- ++(-1.5,0)
			node[midway,above] {\(v\)};
		\node at (0,1) {\(\omega=0\)};
	\end{scope}
	
	\begin{scope}[shift={(10,0)}]
		\node at (0,5) {final state};
		\filldraw[fill=black!20!white,thick] (0,0) arc(-90:270:2);
		\draw[->] (0,2) -- ++(-1,0)
			node[midway,above] {\(v'\)};
		\draw[->] (0,1) arc(-90:0:1)
			node[midway,below,right] {\(\omega'\)};
	
	\end{scope}
	
	\begin{scope}[shift={(16,0)}]
		\node at (0,5) {kinetic friction};
		\filldraw[fill=black!20!white,thick] (0,0) arc(-90:270:2);
		\draw[->] (0,2) -- ++(-1.3,0)
			node[midway,above] {\(v\)};
		\draw[red,->,ultra thick] (0,0) -- ++(1,0)
			node[midway, above] {\(F_\text{f}\)};
	\end{scope}
	
	\begin{scope}[shift={(22,0)}]
		\node at (0,5) {unknown impulse};
		\filldraw[fill=black!20!white,thick] (0,0) arc(-90:270:2);
		\draw[->] (0,2) -- ++(-1.3,0)
			node[midway,above] {\(v\)};
		\draw[blue,->,ultra thick] (0,0) -- ++(-1,0)
			node[midway, above] {\(p\)};
	\end{scope}
\end{tikzpicture}
	\caption{Scenario A: Pushed roller.}
	\label{fig:Bowling}
\end{figure*}
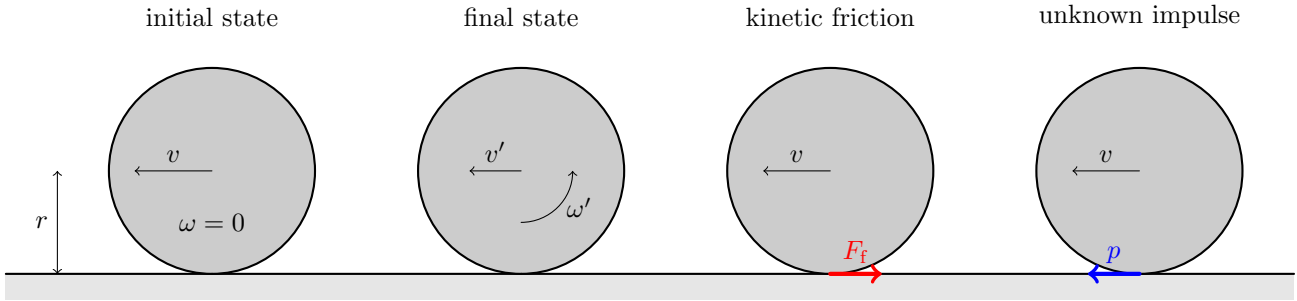

This is arguably the most popular RWS scenario and has been discussed in textbooks~\cite[Example 9-19]{tipler2008} and didactic publications \cite{hierrezuelo1995,desousa1997,deambrosis2015,ozkanlar2020}:
A roller with mass~\(m\), moment of inertia~\(I\) and radius~\(r\) is brought onto a horizontal plane such that it moves with velocity~\(v\) but does not rotate at all, i.e., \(\omega=0\) (see Fig.~\ref{fig:Bowling}).
The kinetic friction between roller and plane depends only on the normal force (and thus is constant).
Rolling friction and air resistance shall be ignored.
The goal is to determine the roller's velocity~\(v'\) when it is not slipping anymore.
Typical examples for this scenario are a bowling ball that has just been thrown onto a lane or a billiard ball that has just been struck.

\subsection{Explicit approach using details of the friction} \label{explicit}
We first describe the typical approach to this problem, employing the constant kinetic friction force \(F_\text{f}\) (see Fig.~\ref{fig:Bowling}).
During the duration~\(t\) of the slipping phase, the roller is subject to a constant negative acceleration~\(a\) and thus its final velocity is:
\begin{equation} \label{Eq:v_1}
	  v'
	= v - at
	= v - \frac{F_\text{f} t}{m}
.\end{equation}
Simultaneously, it is subject to a constant angular acceleration~\(\alpha\) on account of the torque~\(\tau\) caused by~\(F_\text{f}\) and thus its final angular velocity is:
\begin{equation} \label{Eq:omega_1}
	  \omega'
	= \alpha t
	= \frac{\tau t}{I}
	= \frac{r F_\text{f} t}{I}
	\quad\Longrightarrow\quad
	t = \frac{I \omega'}{r F_\text{f}}
\end{equation}
To determine \(v'\), we insert Eq.~\ref{Eq:omega_1} into Eq.~\ref{Eq:v_1} and use that, after the slipping phase, the rolling condition \(v'=\omega' r\) must be met:
\begin{eqnarray} \eqalign{ \label{Eq:solution_1} 
	& v'
	= v - \frac{F_\text{f} t}{m}
	= v - \frac{F_\text{f} I \omega'}{m r F_\text{f}}
	= v - \frac{I v'}{m r^2}\\
	\Longrightarrow\quad &
	  v'
	= \frac{v}{1 + \frac{I}{m r^2}}
	= \frac{m r^2}{m r^2 + I}\, v
. } \end{eqnarray}

This result being independent of the friction force suggests that the details of the friction do not matter and we can treat the transition in a more abstract manner, namely as a collision.

\subsection{Approach using a collision with the ground} \label{Bowling_collision}
We summarise the entirety of the friction forces during the slipping phase into an impulse~\(p\) which the roller exerts on the ground in direction of motion at the point of contact (see Fig.~\ref{fig:Bowling}).
Alternatively, we could work with the opposite impulse that the ground exerts on the roller.
With this, the conservation of (linear) momentum yields:
\begin{equation} \label{Eq:momentum}
	m v = m v' + p
.\end{equation}
Also, we gain from the conservation of angular momentum:
\begin{equation} \label{Eq:angmomentum}
	I \omega' = p r
.\end{equation}
Combining Eqs.~\ref{Eq:angmomentum} and~\ref{Eq:momentum} with the rolling condition \(v'=\omega' r\), we obtain:
\begin{eqnarray} \eqalign{ \label{Eq:solution_2}
	& m v
	= m v' + p
	= m v' + \frac{I \omega'}{r}
	= m v' + \frac{I v'}{r^2}\\
	\Longrightarrow\quad &
	  v'
	= \frac{m v}{m + \frac{I}{r^2}}
	= \frac{m r^2}{mr^2 + I}\, v
, } \end{eqnarray}
which agrees with Eq.~\ref{Eq:solution_1}.

This confirms that the final outcome does not depend on the details of the friction:
It can as well be non-linear or velocity-dependent.
It may be even a totally different interaction with the ground that results in the rolling condition to be assumed.
For instance in the example of the bowling ball, suppose the inside of the grip hole collides with a rigid obstacle in the ground.
Alternatively introduce a small strong demon who is firmly connected to the ground at the initial position of the roller and briefly holds it until the rolling condition is met.

As the last examples are easier to regard as instantaneous, they may serve as an accessible, illustrative, and memorable introduction of our unknown impulse~\(p\).
From there it is only a small step to realising that the interaction can extend over a longer period of time without affecting the outcome.

\subsection{Side note: Approach using an instantaneous axis and Steiner's theorem} \label{Steiner}
There is a slightly different way to solve this problem by summarising:
For a rolling motion without slipping, the roller must perform a rotation around its point of contact with the ground, usually named \emph{instantaneous axis.}
According to Steiner's theorem, the roller's moment of inertia with respect to this axis is \(\check{I} = I + mr^2\).
The conservation of angular momentum around this axis yields:
\begin{eqnarray} \eqalign{ \label{Eq:solution_3}
	& m v r
	= \check{I} \omega'
	= \left ( I+mr^2 \right ) \frac{v'}{r} \\
	\Longrightarrow\quad &
	  v'
	= \frac{m r^2}{mr^2 + I}\, v
. } \end{eqnarray}

While this approach is even more compact than the previous one, we do not consider it as accessible as it relies on the difficult concept of instantaneous axes~\cite{hierrezuelo1995} and the interaction with the ground has become completely implicit.
Also, we consider it to require greater care when applying it to more complex problems such as the scenario described in Sec.~\ref{Monowheel}.
However this approach has its merit by illustrating Steiner's theorem: the rotation around an arbitrary axis can be decomposed into a rotation around the centre of mass and the linear motion of that centre of mass.

\section{Scenario B: Hitting the ground rotating} \label{Hitting}

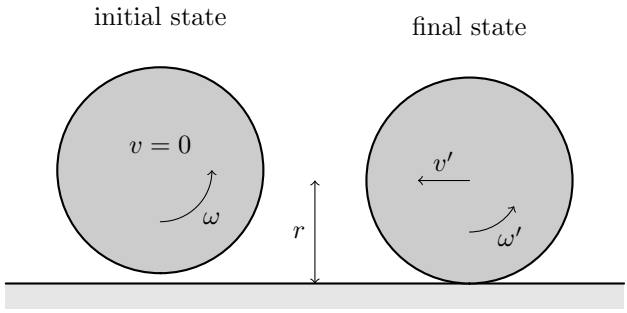
\begin{figure}
	\begin{center}
	\begin{tikzpicture}[scale=0.685,line cap=round]
	\fill[black!10!white]
		(-6,0) rectangle (6,-0.5);
	\draw[thick] (-6,0) -- (6,0);
	
	\draw[<->] (0,0) -- ++(0,2) node[midway,left] {\(r\)};
	
	\begin{scope}[shift={(-3,0.2)}]
		\node at (0,5) {initial state};
		\filldraw[fill=black!20!white,thick] (0,0) arc(-90:270:2);
		\node at (0,2.5) {\(v=0\)};
		\draw[->] (0,1) arc(-90:0:1)
			node[midway,shift={(0.2,-0.15)}] {\(\omega\)};
	\end{scope}
	
	\begin{scope}[shift={(3,0)}]
		\node at (0,5) {final state};
		\filldraw[fill=black!20!white,thick] (0,0) arc(-90:270:2);
		\draw[->] (0,2) -- ++(-1,0)
			node[midway,above] {\(v'\)};
		\draw[->] (0,1) arc(-90:-30:1)
			node[midway,shift={(0.2,-0.15)}] {\(\omega'\)};
	\end{scope}
\end{tikzpicture}
	\end{center}
	\caption{Scenario B: Hitting the ground rotating.}
	\label{fig:Hitting}
\end{figure}

This scenario has also been featured in didactic publications~\cite{suarez2020} and can be regarded the opposite to the previous one:
A roller is released slightly above the ground, while already rotating around its centre with angular velocity~\(\omega\) (see Fig.~\ref{fig:Hitting}).
However, the centre itself does not move, i.e., \(v=0\).
Again, rolling friction and air resistance shall be ignored and the goal is to determine the roller's final velocity~\(v'\).

We again introduce an impulse~\(p\) that the roller exerts on the ground to accelerate.
This time it is opposed to the direction of motion.
The conservation of linear momentum gives us \(mv' = p\) and the conservation of angular momentum gives us:
\begin{equation}
	  I \omega
	= I \omega' + pr
	= \frac{I v'}{r} + m v' r
	~\Longrightarrow~
	v' = \frac{I \omega r}{I + mr^2}
.\end{equation}

The possibility of the roller bouncing can serve to illustrate why the details of the friction do not matter:
One can argue that the roller only is subject to horizontal and rotational acceleration when it is in contact with the ground and thus bouncing phases can be ignored when determining the final state.

\section{Scenario C: Breaking with the monowheel} \label{Monowheel}

\begin{figure}
	\begin{center}
		\newcommand{\driver}{
	\coordinate (shoulders) at (1.1,-0.5);
	\coordinate (hip) at (0.7,-1.27);
	\draw[ultra thick,red,line join=round]
		   (-1.05,-1   )
		-- (-0.9 ,-1.1 )
		-- ( 0.05,-1.0 )
		-- (hip);
	\draw[ultra thick,red,line join=round]
		   (-0.85,-1.02)
		-- (-0.8 ,-1.2 )
		-- (-0.0 ,-1.15)
		-- (hip)
		to[out=40,in=-90] (shoulders)
		-- ++( 0,0.1) arc(-90:270:0.2);
	\draw[ultra thick,red,line join=round]
			(shoulders)
		-- ( 0.6 ,-0.7)
		-- ( 0   ,-0.6);
	\draw[ultra thick,blue,line join=round]
		(0.1,-0.7) -- (-1.1,-0.97)
		(0.1,-1.0) arc [start angle=-90, end angle=270,x radius=0.1, y radius=0.3];
	\draw[ultra thick,red,line join=round]
			(shoulders)
		-- ( 0.65,-0.9)
		-- ( 0.2 ,-0.8);
}

\begin{tikzpicture}[scale=0.685,line cap=round]
	
	\fill[black!10!white]
		(-6,0) rectangle (6,-0.5);
	\draw[thick] (-6,0) -- (6,0);
	
	\draw[<->] (0,0) -- ++(0,2.5) node[midway,left] {\(r\)};
	
	\begin{scope}[shift={(-3,0)}]
		\node at (0,6) {before breaking};
		\filldraw[fill=black!20!white,thick]
			(0,0) arc(-90:270:2.5)
			(0,1) arc(270:-90:1.5);
		\begin{scope}[shift={(0,2.5)}]
			\driver
		\end{scope};
		\node at (0,3) {\(\omega_\text{i} = 0\)};
		
		\draw[<-]
			(-1.0,0.768)
			node[right,yshift=-4] {\(\omega_\text{w}\)}
			arc(-120:-170:2);
		
		\draw[->]
			(0.8,4.6) -- ++(-1.6,0) node[midway,below] {\(v\)};
	\end{scope}
	
	\begin{scope}[shift={(3,0)}]
		\node at (0,6) {after breaking};
		\filldraw[fill=black!20!white,thick]
			(0,0) arc(-90:270:2.5)
			(0,1) arc(270:-90:1.5);
		\begin{scope}[shift={(0,2.5)},rotate=120]
			\driver
		\end{scope};
		
		\draw[<-]
			(-0.5,1.634)
			node[right] {\(\omega_\text{i}'\)}
			arc(-120:-150:1);
		\draw[<-]
			(-1.0,0.768)
			node[right,yshift=-4] {\(\omega_\text{w}'=\omega_\text{i}'\)}
			arc(-120:-150:2);
		
		\draw[->]
			(0.5,4.7) -- ++(-1,0) node[midway,below] {\(v'\)};
	\end{scope}
\end{tikzpicture}
	\end{center}
	\caption{Scenario C: Breaking with the monowheel.}
	\label{fig:Monowheel}
\end{figure}
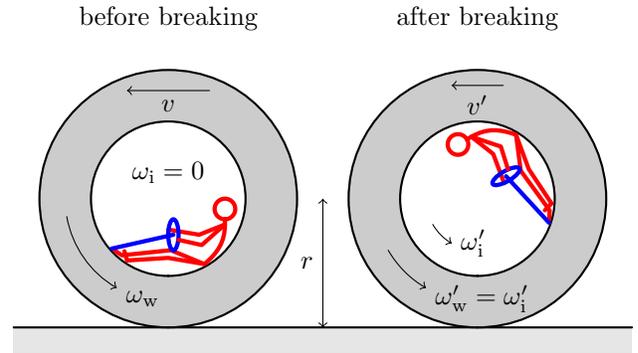

A monowheel~\cite{cardini2006} is an exotic vehicle, which consists of two parts (see Fig.~\ref{fig:Monowheel}):
a giant wheel~(w) and an \emph{interior}~(i) containing the driver, motor, etc.
The interior and wheel can be turned against each other by motor or muscle power to make the vehicle roll.
During regular operation, the wheel rolls, but the interior barely rotates (\(\omega_\text{i}=0\)).
For those unfamiliar with monowheels, we recommend to watch them on video, e.g., Refs.~\citeonline{pathe1932} and~\citeonline{bi2018}.

We here consider the scenario of a monowheel driving with a velocity~\(v\).
The vehicle has a mass~\(m\) and radius~\(r\).
The wheel's moment of inertia is~\(I_\text{w}\) and that of the interior is~\(I_\text{i}\), both with respect to the centre.
Now, the monowheel makes a \emph{hard brake,} i.e., wheel and interior are locked afterwards.
It briefly skids and then continues rolling.
The goal is to determine the final velocity~\(v'\) after this.
Once more, rolling friction and air resistance shall be ignored.

In this scenario, two problems are combined:
Hitting the breaks leads to a rotational collision~\cite{Bougie2019, Danna2021} of the two components of the monowheel, which causes a mismatch in the rolling condition and thus a rolling-with-slipping problem.
Relatedly, there are two points of friction: between interior and wheel as well as between wheel and ground.
Still, our approach allows to handle both at once:

We first note that the rolling condition is met before and after the breaking, i.e., \(v=\omega r\) and \(v' = \omega' r\).
With an impulse~\(p\) exerted on the ground by the monowheel, we obtain for the conservation of momentum:
\begin{equation}\label{Eq:mw_momentum}
	mv = mv' + p
,\end{equation}
and for the conservation of angular momentum:
\begin{eqnarray} \eqalign{\label{Eq:mw_angular}
	& I_\text{w} \omega
	= \left( I_\text{w} + I_\text{i} \right) \omega' - pr \\
	\Longrightarrow\quad &
	  \frac{ I_\text{w} v }{r}
	= \frac{\left( I_\text{w} + I_\text{i} \right) v'}{r} - \left( mv - mv' \right) r \\
	\Longrightarrow\quad &
	  I_\text{w} v
	= \left( I_\text{w} + I_\text{i} \right) v' - mvr^2 - mv'r^2 \\
	\Longrightarrow\quad &
	  v'
	= \frac{ mr^2 + I_\text{w} }{ mr^2 + I_\text{w} + I_\text{i} }\, v
.}\end{eqnarray}

\section{Conclusion}

We demonstrated how to determine the final state in problems involving rolling with slipping (RWS) by introducing a momentum transferred to the ground.
In contrast to the usual approach~\cite{hierrezuelo1995, desousa1997, pinto2001, tipler2008, maidana2016, suarez2020, ozkanlar2020, kajiyama2021}, we did not need to calculate frictional forces and thus could show that thresults are more general.
Moreover, we showed that our approach allowed to solve a complex problem involving two frictional forces.

We propose to use our approach to demonstrate how conservation laws and treating complex events as collisions may ease solving problems, in particular in the context of rotations.
Moreover it may be illustrative to contrast our approach with more explicit ones or ones using instantaneous axis and Steiner's theorem (Sec.~\ref{Steiner}) or thermodynamics~\cite{desousa1997}.
While there are several existing examples which can be explored in two different ways with the one using conservation laws being the easier one, they mostly use energy conservation.
We therefore consider our approach a valuable and complementary alternative for teaching these concepts.

\appendix
\section{Exercises}
We here provide a sequence of two exercises surrounding the presented technique using the formerly discussed scenarios.
TikZ sources of all pictures are provided as supplement for easy re-use and adaption.

In the first exercise, the students are first guided along the explicit approach of Sec.~\ref{explicit} and are then introduced to the alternative presented in Sec.~\ref{Bowling_collision}.
In the second exercise, the students are assumed to be familiar with the technique and are challenged to apply it to the problem of Sec.~\ref{Monowheel}, which is embedded in other tasks revolving around the monowheel.
Both exercises are designed to be solved in groups.
A small exercise on the problem discussed in Sec.~\ref{Hitting} can be inserted in between.

\subsection{Exercise 1: Bowling ball}
You throw a bowling ball on a lane.
Immediately afterwards, it slides with a velocity \(v=\unitfrac[5]{m}{s}\) but it does not rotate.
Afterwards, kinetic friction causes it to rotate until it rolls losslessly with a velocity~\(v'\).
The goal of this exercise is to compute the final velocity~\(v'\) in two different ways.
Assume the bowling ball to be a perfect homogeneous sphere with radius~\(r=\unit[10]{cm}\) and weight \(m=\unit[5]{kg}\).
The coefficient of kinetic friction between bowling ball and ground is \(\mu=0.2\).

\begin{enumerate}
	\item
		Compute the friction force~\(F_\text{f}\) exerted on the bowling ball and the resulting acceleration~\(A\) as well as the angular acceleration~\(\alpha\).
	\item
		After what time~\(T\) does the bowling ball meet the rolling condition?
	\item
		What is the ball's final velocity~\(v'\)?
	\item
		\emph{The results of the previous tasks showed that the transition happens quickly and that the result is independent of \(\mu\).
		This suggests that we can treat the entire process as a collision.}
		
		Introduce an unknown momentum~\(p\) that summarises the interaction of the ball with the ground.
		It acts on the point of contact and in direction of rolling.
		Then use this impulse in the conservation of momentum and angular momentum and solve the resulting equations for~\(v'\).
\end{enumerate}

\subsection{Exercise 2: Breaking with the monowheel}
Watch the videos in Refs.~\citeonline{pathe1932} or~\citeonline{bi2018} to familiarise yourself with the monowheel.
Now consider a monowheel which consists of two parts:
\begin{itemize}
	\item
		The wheel, which is a homogeneous cylindrical shell, with inner radius~\(r_2=\unit[145]{cm}\), outer radius~\(r_1=\unit[150]{cm}\), width~\(b=\unit[100]{cm}\) and density \(\rho_1 = \unitfrac[4]{g}{cm^3}\).
	\item
		An interior containing the driver, motor, etc., which is simplified as a homogeneous quarter cylinder, with radius~\(r_2=\unit[145]{cm}\), width~\(b\) and density \(\rho_1 = \unitfrac[0.5]{g}{cm^3}\).
\end{itemize}

\begin{enumerate}
	\item
		Determine the wheel's moment of inertia~\(I_\text{w}\) and that of the interior~\(I_\text{i}\), both with respect to the wheel's centre.
	\item
		When driving with a velocity \(v=\unitfrac[30]{km}{h},\) the monowheel makes a ``hard brake'', such that the interior and wheel move as one.
		Show that the velocity after breaking is:
		\[
			v'
			= \frac{ mr_1^2 + I_\text{w} }{ mr_1^2 + I_\text{w} + I_\text{i} }\, v
		,\]
		where \(m\) is the mass of the vehicle.
		Compute~\(v'\).
	\item
		You want to design a monowheel that optimises the reduction of velocity (\(\frac{v'}{v}\)) when making a hard break.
		You can arbitrarily distribute the monowheel's mass to wheel and interior and also arbitrarily position it with respect to the centre.
		(The interior must not be outside the wheel though.)
		What is the lowest \(\frac{v'}{v}\) you can achieve?
	\item
		Is there a better way to halt a monowheel?
		Why is this a technical and safety challenge?
		Compare with a car.
\end{enumerate}

\providecommand{\newblock}{}

\end{document}